# A Full Performance Analysis Of Channel Estimation Methods for Time Varying OFDM Systems

Aida Zaier[1] and Ridha Bouallègue[2]

[1]National Engineering School of Tunis, Tunis University, Tunisia,
zaieraida@yahoo.fr

[2] High School of Communications,Tunis Ariana, Tunisia
ridha.bouallegue@gnet.tn

## ABSTRACT

*In this paper, we have evaluated various methods of time-frequency-selective fading channels estimation in OFDM system and some of them improved under time varying conditions. So, these different techniques will be studied through different algorithms and for different schemes of modulations (16 QAM, BPSK, QPSK,…).*

*Channel estimation gathers different schemes and algorithms, some of them are dedicated for slowly time varying (such as block type arrangement insertion, Bayesian Cramer-Rao Bound, Kalman estimator, Subspace estimator,…) whereas the others concern highly time varying channels (comb type insertion,…) . There are others methods that are just suitable for stationary channels like blind or semi blind estimators.
For this aim, diverse algorithms were used for these schemes such as Least Squares estimator LS, Least Minimum Squares LMS, Minimum Mean-Square-ErrorMMSE, Linear Minimum Mean-Square-Error LMMSE, Maximum Likelihood ML,…to refine estimators shown previously.*

## 1. INTRODUCTION

The increasing require for high-bit-rate digital mobile communications has incited the appearance of Orthogonal Frequency-Division Multiplexing (OFDM) for achieving good performance in high rate data transmission. It is also an effective technique that produces a high spectral efficiency and a good scheme to combat frequency-selective fading channels in wireless communication systems without forgetting the major property that is subcarrier orthogonality.

Hence, the symbol duration must be significantly larger than the channel delay spread. In orthogonal frequency  division multiplexing (OFDM) [1][2], the entire channel is divided into many narrow subchannels. Splitting the high-rate serial data stream into many low-rate parallel streams, each parallel stream modulates orthogonal subcarriers by means of the inverse fast Fourier transform (IFFT).[3]

If the bandwidth of each subcarrier is much less than the channel coherence bandwidth, a frequency flat channel model can be assumed for each subcarrier. Moreover, inserting a cyclic prefix (or guard interval) results in an inter-symbol interference (ISI) free channel assuming that the length of the guard interval is greater than the delay spread of the channel. Therefore, the effect of the multipath channel on each subcarrier can be represented by a single complex multiplier, affecting the amplitude and phase of each subcarrier. Hence, the equalizer at the receiver can be implemented by a set of complex multipliers, one for each subcarrier. [3]

Under multi path spread situation, a dynamic estimation of channel is necessary before the demodulation of OFDM signals to ensure a coherent detection and since the radio channel is frequency selective and time-varying for wideband mobile communication systems [4][5].

In the literature, many channel estimation schemes are found and depends on if the channel is constant, slowly or fast time varying.

Traditionally, channel estimation is achieved by sending training sequences through the channel. However, when the channel is varying, even slowly, the training sequence needs to be sent periodically in order to update the channel estimates. Hence, the transmission efficiency is reduced. The increasing demand for high-bit-rate digital mobile communications makes blindchannel identification and equalization very attractive, since they do not require the transmission of a training sequence. Early methods for blind channel equalization exploit higher-order statistics (H0S) of the outputs that are sampled at symbol rate. [6]

Heath and Giannakis[8]proposed a subspace method using cyclic correlation of the channel output to blindly estimate the channel in OFDM systems. But the estimated channel error in that study is large a subspace approach based on second order statistics is proposed to blindly identify the channel in OFDM systems. The channel identifiability is due to the cyclostationarity inherent in the OFDM systems with cyclic prefix. We derive a sufficient condition that guarantees all the channels to be identifiable no matter what their zero locations are. The difference between the proposed algorithm in this paper and the method in [7] is **as** follows. The approach in [7] uses the cyclic correlation that is defined **as** the Fourier series expansion of the time-varying correlation of the received data samples. In [6] , the time-invariant autocorrlation of the vector that consists of *N* blocks of the received data was used.

For fast time-varying channels, many existing works resort to estimating the equivalent discrete-time channel taps, which are modeled by the basis expansion model (BEM) [9] [10]. The BEM methods [9] are Karhunen-Loeve BEM (KL-BEM), prolate spheroidal BEM (PS-BEM), complex exponential BEM (CE-BEM) and polynomial BEM (P-BEM). The KL-BEM is optimal in terms of mean square error (MSE), but is not robust to statistical channel mismatches, whereas the PS-BEM is a general approximation for all kinds of channel statistics, although its band-limited orthogonal spheroidal functions have maximal time concentration within the considered interval. The CE-BEM is independent of channel statistics, but induces a large modeling error. Finally, a great deal of attention has been paid to the P-BEM [10], although its modeling performance is rather sensitive to the Doppler spread; nevertheless, it provides a better fit for low, than for high Doppler spreads. In [3], a piece-wise linear method is used to approximate the channel taps, and the channel tap slopes are estimated from the cyclic prefix or from both adjacent OFDM symbols. [11]

Several others methods was investigated to estimate channel by using the Cramer-Rao Bounds (CRBs), which give fundamental lower limits of the Mean Square Error (MSE) achievable by any unbiased estimator. A Modified CRB (MCRB), easier to evaluate than the Standard CRB (SCRB), has been introduced in [12] [13]. The MCRB proves useful when, in addition to the parameter to be estimated, the observed data also depend on other unwanted parameters. More recently, the problem of deriving CRBs, suited to time-varying parameters, has been addressed throughout the Bayesian context. In [9], the authors propose a general framework for deriving analytical expression of on-line CRBs. [14]

## 2. SYSTEM MODEL

The baseband OFDM system is practically the same for all the schemes of channel estimations and differs just from the bloc of the channel but some schemes can add another bloc used especially for interpolation or for equalization,…
Then we will present in this section the major systems used by the majority of channel estimation schemes.

### 2.1. OFDM System for Channel Estimation based on Pilot Arrangement

The OFDM system model used for training sequence (pilot signal) consists of mainly of a mapper bloc forwarded by a S/P conversion, then there is an insertion pilot bloc follow up of an IDFT calculation of the information data. After that we find the guard insertion bloc and a P/S conversion before reaching the channel which is affected by an AWGN noise. The data stream will be converted on a parallel stream, then the guard interval is removed and we will sail towards the frequency domain. Channel estimation is afterward performed before carrying out a P/S conversion and attainment of the demapper bloc to restore back the data stream.

We will give here an overview about the routing circuit of the transmission and estimation process according to pilot arrangement channel estimation.
Assuming first that a 16 QAM modulation will be used here, the input binary data is mapped according to this modulation scheme. After crossing the S/P bloc, the pilot used here will be inserted in all sub carriers of one OFDM symbol with a specific period or uniformly between data sequence for a block pilot type estimator or in some specific sub carriers for the comb pilot type.
Then, the data sequence will pass up the IDFT block for the transformation to time domain and the expression of x(n) (N being the DFT length) is given as follow:

$$x(n) = IDFT\{X(k)\} \quad n = 0,1,2,\ldots,N-1$$
$$= \sum_{k=0}^{N-1} X(k)\, e^{j(2\pi kn/N)} \quad (1)$$

After that, the cyclic prefix will be inserted to preserve orthogonality of the sub carriers on the one hand and to evade inter symbol interference between adjacent OFDM symbols on the other hand. The guard time which contains this cyclic prefix having a length $N_g$ is chosen to be grater than the delay spread. Then the resulting symbol is:

$$x_f(n) = \begin{cases} x(N+n), & n = -N_g, -N_g+1, \ldots, -1 \\ x(n), & n = 0,1,\ldots,N-1 \end{cases} \quad (2)$$

After a P/S conversion, the OFDM symbol will cross the channel expected to be frequency selective and time varying fading channel with an additive noise and will be given by:

$$y_f(n) = x_f(n) \otimes h(n) + w(n) \tag{3}$$

h(n) is the channel impulse response represented by [16]

$$h(n) = \sum_{i=0}^{r-1} h_i e^{j(\frac{2\pi}{N})f_{Di}T_n} \delta(\lambda - \tau_i), \quad 0 \leq n \leq N-1 \tag{4}$$

Where r is the total number of propagation paths, $h_i$ is the complex impulse response of the *i*th path, $f_{Di}$ is the *i*th path Doppler frequency shift, $\lambda$ is the delay spread index, T is the sample period and $\tau_i$ is the *i*th path delay normalized by the sampling time. [17]
The guard interval is removed after the return to continuous domain from the S/P block and the resulting signal is:

$$y_f(n) \; for - N_g \leq n \leq N-1$$

$$y(n) = y_f(n + N_g) \; n = 0,1,2,\ldots,N-1 \tag{5}$$

The frequency form of this resulting signal will be expressed as follow:

$$Y(k) = DFT\{y(n)\} \; k = 0,1,2,\ldots,N-1$$
$$= \frac{1}{N}\sum_{n=0}^{N-1} y(n) e^{-j(\frac{2\pi k n}{N})} \tag{6}$$

By supposing a transmission without an inter symbol interference ISI, the relation between the frequency components is [18]:

$$Y(k) = X(k)H(k) + I(k) + W(k) \tag{7}$$

Where $H(k) = DFT\{h(n)\}$, $I(k)$ is the inter-carrier interference because of the Doppler frequency and $W(k) = DFT\{w(n)\}$
The pilot signals are then extracted and cross the channel estimation block, after that the estimated channel H_e(k) for the data sub-channel is obtained and the transmitted data is estimated. At last, the binary data sequence is recovered by the signal demapper block.

### 2.2. OFDM System for Blind Estimation

In OFDM systems, the serial data are converted into M parallel streams. Each parallel data stream modulates a different carrier. The frequency separation between the adjacent carriers is 1/T, where *T* is the symbol duration for the parallel data that is *M* times of the symbol duration for the serial data. Let us consider an OFDM signal in the interval (nT,(n+1)T) as [6]:

$$s(t) = \sum_{m=0}^{M-1} a_m(n) e^{jw_m t}$$

$$\tag{8}$$

Where $a_m(n)$ are symbols resulting from a modulation constellation like 16 QAM. $w_m$ is the frequency of mthcarrier that is $m\frac{2\pi}{T}$. The M samples that are sampled at $t = nT + i\frac{1}{T}, i = 0,1,\ldots, M-1$ are as follow:

$$s(nM + i) = \sum_{m=0}^{M-1} a_m(n) e^{j\frac{2\pi}{M}mi} \quad (9)$$

From this equation, the M samples can be seen as the inverse discrete Fourier transform (IDFT) of a block for M input symbols.

Theoretically speaking, when the number of carriers is large enough, symbol duration $T$ is much larger than the duration of FIR channel; IS1 is negligible. However, for the high-bit-rate communications, it is impractical to choose very large M to make ISI negligible. Therefore, a cyclic prefix of length P is added into each block of IDFT output at the transmitter. The length of the prefix is chosen to be longer than the length of the channel impulse response in order to avoid inter-block interference (IBI). That results with total cancellation of IS1 and inter carrier interference (ICI). The input data will be as follow:

$$s(n(M + P) + i) = \sum_{m=0}^{M-1} a_m(n) e^{j\frac{2\pi}{M}(i-P)}$$
$$i = 0,1,\ldots, M + P - 1$$

(10)

Where $s(n(M + P) + i), i = 0,1,\ldots, P - 1$ denotes the cyclic prefix.

Assuming that the channel is frequency selective affected by an additive white Gaussian noise (AWGN), The received signal r(n) will be degraded by theses two conditions. We will suppose also that the length $L$ of the channel impulse response is known. Assuming that blocks are synchronized and carrier frequency offset is corrected [19], the receiver removes the first $P$ symbols corresponding to the cyclic prefix and performs an M-point DFT on the remaining samples of received signal to obtain $y_i(n), i = 0,1,\ldots, M - 1$. If the cyclic prefix duration is equal or more than the channel duration, i.e. $P \geq L$, it is shown that [20]

(11)

$$y_i(n) = a_i(n) H\left(\frac{2\pi}{M} i\right) + v_i(n)$$

Where H(.) is the frequency response of the channel. It is evident from the equation above that the ISI is completely cancelled and the effect of the channel at the receiver is simply a complex gain and AWGN

### 2.3. OFDM System for Channel Estimation based on Kalman Filter

For this scheme of estimation, the OFDM system model used is a standard one in which the information symbols are grouped into blocks and inverse discrete Fourier transform (IDFT) is performed on each block and cyclic prefix (CP) added before they are fed into the modulator and transmitted. At the receiver, DFT is performed on each received OFDM symbol after the CP is removed. With proper CP extensions, carrier synchronization and sample timing of tolerable leakage, the sample from the kth subcarrier for the nth OFDM symbol is [21]:

$$y_k(n) = H_k(n) s_k(n) + w_k(n), k = 1, \ldots, N, -\infty < N < \infty \quad (12)$$

Where $s_k(n)$ is the kth information symbol of the nth OFDM symbol, $H_k(n)$ is the gain of the kth sub-channel during the nth OFDM symbol, $w_k(n)$ is the noise, and N is the total number of the subcarriers. In this study, we assume that $s_k(n)$ is drawn from a BPSK constellation $\{-1, +1\}$ independently for different k and n, and $w_k(n)$ is a circular Gaussian random variable with zero mean and variance $\sigma_n^2$ and i.i.d for different k and n. We also assume that the channel is Rayleigh fading and we use Jakes' model [22] for the power spectral density and Doppler spectrum of the fading process. Specifically, we have the correlation of the channel gain $H_k(n)$ as

$$r_{k,l}(m) = E\{H_k(n) H_l^*(n-m)\} \quad (13)$$

$$= J_0(2\pi f_d mT) \frac{1 - j2\pi(l-k)\sigma_t/T}{1 + 4\pi^2 (l-k)^2 \sigma_t^2/T^2}$$

where $f_d$ is the maximum Doppler frequency, $\sigma_t$ the maximum delay spread of the channel and T the OFDM symbol duration, and $J_0(.)$ is the zeroth-order Bessel function of the first kind.[21]
In general, the dynamics of the gain $H_k(n)$ can be well modeled by an autoregressive model AR. Defining h[n] as $h[n] := [H_1(n), \ldots, H_N(n)]^T$, the kth order AR model for h[n] can be written as:

$$h[n] = -\sum_{i=1}^{k} A[i] h[n-i] + Q u[n] \quad (14)$$

Where $A[1], \ldots, A[k]$ and $Q$ are N*N matrices and $u[n]$ is a N*1 vector white Gaussian process. $A[1], \ldots, A[k]$ and $Q$ are the model parameters which are obtained by solving a Yule Walker equation using the expression of $r_{k,l}(m)$.
Based on the AR model of the channel, a state-space model for the OFDM system can be built. By defining $x[n] := [h^T[n], \ldots, h^T[n-k+1]]^T$, we write:

$$x[n] = Cx[n-1] + Gv[n]$$

$$(15)$$

The equation above can be seen as the state equation with $v[n]$ being an AWGN noise. C and G are given by:

$$C = \begin{bmatrix} -A[1] & -A[2] \cdots & -A[k] \\ I_N & 0_N \cdots & 0_N \\ 0_N \cdots & \ddots\, I_N & \ddots\, 0_N \end{bmatrix}$$

$$G := [Q, 0_K, \ldots, 0_K]^T$$

$I_N$ and $0_N$ are respectively N*N identity matrix and all zeros matrix. The observation equation of the state space model is a vector version of y[n]

$$y[n] = D[n]x[n] + w[n] \quad (16)$$

Where $y[n] := [y_1[n], \ldots, y_N[n]]^T$, $D := [S[n], 0_N, \ldots, 0_N]^T$, $w[n] := [w_1[n], \ldots, w_N[n]]^T$ and $S[n]$ is an N*N diagonal matrix with $s_k[n]$ being its kth diagonal entry.

## 2.4. OFDM System Model based Bayesian Cramer-Rao Bounds (BCRBS)

The model adopted in this case is very near to the classical system model that is consists of N sub-carriers and a cyclic prefix of length $N_g$. Supposing that $T_s$ denotes the sampling time and $v = N + N_g$, the duration of an OFDM symbol will be $T = vT_s$.

In the case of a transmission over a multipath Rayleigh channel and for a *n*th transmitted OFDM symbol $x_{(n)}$ given as $x_{(n)} = [x_{(n)}\left[-\frac{N}{2}\right], x_{(n)}\left[-\frac{N}{2}+1\right], \ldots, x_{(n)}\left[\frac{N}{2}-1\right]]^T$ ({$(x_{(n)}[b]$} are modulated 4-QAM or 16QAM ), the *n*th received OFDM symbol $y_{(n)} = [y_{(n)}\left[-\frac{N}{2}\right], y_{(n)}\left[-\frac{N}{2}+1\right], \ldots, y_{(n)}\left[\frac{N}{2}-1\right]]^T$ is as follow: [14]

$$y_{(n)} = H_{(n)} x_{(n)} + w_{(n)} \tag{17}$$

where $w_{(n)}$ is a $N \times 1$ zero-mean complex Gaussian noise vector with covariance matrix $\sigma^2 I_N$, and $H_{(n)}$ is a $N \times N$ diagonal matrix with diagonal elements given by:

$$[H_{(n)}]_{k,k} = \frac{1}{N} \sum_{l=1}^{L} [\alpha_l^{(n)} \times e^{-j2\pi(\frac{k-1}{N}-\frac{1}{2})\tau_l}] \tag{18}$$

$L$ is the total number of propagation paths, $\alpha_l$ is the *l*th complex gain of variance $\sigma_{\alpha l}^2$ with $\sum_{l=1}^{L} \sigma_{\alpha l}^2 = 1$ and $\tau_l \times T_s$ is the lth delay ($\tau_l < N_g$)

The $L$ individual elements of $\{\alpha_l^{(n)}\}$ are uncorrelated with respect to each other. They are wide-sense stationary narrowband complex Gaussian processes, with the so-called Jakes' power spectrum [23] with Doppler frequency $f_d$. It means that $\alpha_l^{(n)}$ are correlated complex gaussian variables with zero -means and correlation coefficients given by:

$$R_{\alpha l}^k = E\left[\alpha_l^{(n)} \alpha_l^{(n-p)H}\right] = \sigma_{\alpha l}^2 J_0(2\pi f_d Tk) \tag{19}$$

Then, the observation model for the *n*th OFDM symbol can be re-written as:
$$y_{(n)} = diag\{x_{(n)}\} F \alpha_{(n)} + w_{(n)} \tag{20}$$

Where $\alpha_{(n)} = [\alpha_1^{(n)}, \ldots, \alpha_L^{(n)}]^T$ is a $L \times 1$ vector and F is the $N \times L$ Fourier matrix defined by:[14]

## 3. CHANNEL ESTIMATION ALGORITHMS

All the channel estimation schemes presented in the previous section are achieved by using the standard estimators such as Least Squares algorithm (LS), Minimum Mean Square Error (MMSE), Linear Minimum Mean Square Error (LMMSE), Low rank LMMSE, Maximum Likelihood (ML),…

Thus we will present here the majority if these algorithms and scow in the next section their behavior and their impact on channel estimations.

### 3.1 Least Square Estimator:

The LS estimator is shown to be the basic algorithm and gives regular results. Used with practically all the schemes of channel estimation, the LS estimator will be expressed as a ratio between the input data sequence and the output.
Denote X as the input data sequence and Y as the output one, then the LS estimator of the channel attenuation vector can be simply expressed as:

$$\hat{h}_{ls} = X^{-1}y = \left[\frac{y_0}{x_0} \frac{y_1}{x_1} \ldots \frac{y_{N-1}}{x_{N-1}}\right]^T \quad (22)$$

### 3.2. Least Mean Square Estimator:

The LMS estimator uses one tap LMS adaptive filter at each pilot frequency. The first value is found directly through LS and the following values are calculated based on the previous estimation and the current channel output.

The LMS estimator is used mainly for the tracking of the channel and is usually clustered with an equalizer or a decision feedback equalizer.

### 3.3. Minimum Mean Square Error Estimator::

Inter symbol interference ISI is supposed usually dropped by the guard interval, thus the equation of Y [17] will be written as:

$$Y = XFh + W \quad (23)$$

Where

$$X = diag\{X(0), X(1), \ldots, X(N-1)\}$$

$$Y = diag\{Y(0), Y(1), \ldots, Y(N-1)\}^T$$

$$W = diag\{W(0), W(1), \ldots, W(N-1)\}^T$$

$$H = diag\{H(0), H(1), \ldots, H(N-1)\}^T = DFT_N(h)$$

$$F = \begin{bmatrix} W_N^{00} & \cdots & W_N^{0(N-1)} \\ \vdots & \ddots & \vdots \\ W_N^{(N-1)0} & \cdots & W_N^{(N-1)(N-1)} \end{bmatrix}$$

$$W_N^{nk} = \frac{1}{N} e^{-j2\pi\left(\frac{n}{N}\right)k} \quad (24)$$

Then, the MMSE estimate of h is given by [17]:

$$H_{MMSE} = FR_{hY}R_{YY}^{-1}Y \qquad (25)$$

Where

$$R_{hY} = E\{hY\} = R_{hh}F^H X^H \qquad (26)$$

$$R_{YY} = E\{YY\} = XFR_{hh}F^H X^H + \sigma^2 I_N \qquad (27)$$

design respectively the cross covariance matrix between h and Y and the auto- covariance matrix of Y. $R_{YY}$ is the auto-covariance matrix of h and $\sigma^2$ is the noise variance $E\{|W(k)|^2\}$.

### 3.4. LMMSE Estimator:

The LMMSE estimate of the channel attenuations h from the received data **y** and the transmitted symbols X is [17][24]

$$\hat{h}_{lmmse} = R_{hh_{ls}} R_{h_{ls}h_{ls}}^{-1} \hat{h}_{ls}$$

$$= R_{hh}(R_{hh} + \sigma_n^2((XX^h)^{-1})^{-1} \hat{h}_{ls} \qquad (28)$$

Where

$$\hat{h}_{ls} = X^{-1}y = \left[\frac{y_0}{x_0} \frac{y_1}{x_1} \cdots \frac{y_{N-1}}{x_{N-1}}\right]^T \qquad (29)$$

is the least-squares (LS) estimate of h, $\sigma_n^2$ is the variance of the additive channel noise, and the covariance matrices are

$$R_{hh} = E\{hh\}^H \qquad (30)$$

$$R_{hh_{ls}} = E\{h\hat{h}_{ls}\}^H \qquad (31)$$

$$R_{h_{ls}h_{ls}} = E\{\hat{h}_{ls}\hat{h}_{ls}^H\} \qquad (32)$$

The LMMSE estimator (24) is of considerable complexity, since a matrix inversion is needed every time the training data in X changes. We reduce the complexity of this estimator by averaging over the transmitted data [15], *i.e.* we replace the term $(XX^h)^{-1}$ in (27) with its expectation

$E(XX^h)^{-1}$. Assuming the same signal constellation on all tones and equal probability on all constellation points, then we get $E(XX^h)^{-1} = E|1/x_k|^2 I$, here I is the identity matrix. Defining the average signal-to-noise ratio as

$$SNR = E|x_k^2|/\sigma_n^2 \qquad (33)$$

, we obtain a simplified estimator

$$\hat{h} = R_{hh}\left(R_{hh} + \frac{\beta}{SNR}I\right)^{-1}\hat{h}_{ls} \tag{34}$$

Where $\beta = E|x_k^2|E\left|\frac{1}{x_k^2}\right|^2$ is a constant depending on the signal constellation. In the case of 16-QAM transmission, $\beta = 17/9$. Because X is no longer a factor in the matrix calculation, no inversion is needed when the transmitted data in X changes. Furthermore, if $R_{hh}$ and SNR are known before hand or are set to fixed nominal values, the matrix $R_{hh}\left(R_{hh} + \frac{\beta}{SNR}I\right)^{-1}$ needs to be calculated only once. Under these conditions the estimation requires $N$ multiplications per tone.

### 3.5. Optimal Low - rank Estimator

This estimator is generally used to overcome the high computational complexity of the LMMSE algorithm

The optimal rank reduction of the estimator in (30), using the singular value decomposition (SVD), is obtained by exclusion of base vectors corresponding to the smallest singular values [25]. We denote the SVD of the channel correlation matrix

$$R_{hh} = U\Lambda U^H \tag{35}$$

where U is a matrix with orthonormal columns $u_0, u_1, \ldots, u_{N-1}$ and $\Lambda$ is a diagonal matrix, containing the singular values $\lambda_0 \geq \lambda_1 \geq \cdots \geq \lambda_{N-1} \geq 0$ on its diagonal. This allows the estimator in (18) to be written

$$\hat{h} = U\Delta U^H\hat{h}_{ls} \tag{36}$$

where $\Delta$ is a diagonal matrix containing the values $\delta_k = \frac{\lambda_k}{\lambda_k + \frac{\beta}{SNR}} k = 0,1,\ldots,N-1$ on its diagonal. The best rank -$p$ approximation of the estimator in (18) then becomes

$$\hat{h}_p = U\begin{pmatrix}\Delta_p & 0\\ 0 & 0\end{pmatrix}U^H\hat{h}_{ls} \tag{37}$$

Where $\Delta_p$ is the upper left $p\times p$ corner of $\Delta$.
Viewing the unitary matrix $U^H$ as a transform y, the singular value $\lambda_k$ of $R_{hh}$ is the channel energy contained in the $k$th transform coefficient after transforming the LS estimate $\hat{h}_{ls}$. [17].

### 3.6. Maximum Likelihood Estimator

This estimator is well adapted for slowly time varying channel.
Considering an OFDM system model with $N_c$ carriers and M OFDM symbols. The input data X is expressed as: $X = [x_0 \ldots x_{M-1}]$ where $x_m = [x_{0,m}, \ldots, x_{N_c-1,m}]^T$ is an OFDM symbol. For a single OFDM symbol, the received signal is:

$$y = x \odot H + n \quad (38)$$

Where $\odot$ denotes the Hadamard product of the columns of $x$ with H.
After IFFT and CP insertion, the transmitted signal is $X_m = [X_{N_c-N_p,m} \ldots X_{N_c-1,m} X_{0,m} \ldots X_{N_c-1,m}]^T$ where $N_p$ is the size of the prefix and $X = [X_0 \ldots X_{M-1}]$. [25]
After prefix removal and DFT computing, the received signal is given as follow:

$$y_m = x_m \odot H + n \quad (39)$$

Where h designs the channel $h = [h_0 \ldots h_{N_h-1}]$ with $N_h \leq N_p$ and H is the FFT of h; $H = [H_0 \ldots H_{N_c-1}]^T$.

As (34) shows, the OFDM system can be described as a set of parallel Gaussian channels. Because the time-domain channel has a finite length (smaller than the prefix length in a welldesigned OFDM system), these parallel channels feature correlated attenuations. Considering, without loss of generality $x = [1\ 1 \ldots 1]^T$, y will be written as [25]:

$$y = F \begin{bmatrix} h \\ 0 \end{bmatrix} + n = H + n \quad (40)$$

Where $F$ is a $N_c \times N_c$ FFT matrix. The vector $y$ is a Gaussian random variable with mean $F[h\ 0]^T$ and covariance $C_{nn}$. However, the signal part of y is contained only in the space spanned by its mean. Separating the "signal subspace" from the "noise-only subspace," the received signal can be rewritten as

$$y = [F_h F_n] \begin{bmatrix} h \\ 0 \end{bmatrix} + n \quad (41)$$

The reduced signal space is then drawn by:
$$r = F_h^\# y = h + F_h^\# n = h + v \quad (42)$$

Where $v$ is a zero mean Gaussian noise of covariance $C_{vv} = F_h^H C_{nn} F_h$ and $F_h^\# = (F_h^H F_h)^{-1} F_h^H$. If $C_{nn} = \sigma_n^2 I_{N_c}$, $v$ is a white Gaussian noise of covariance matrix $\sigma_n^2 I_{N_h}$. The reduced space (Gaussian) signal has a log-likelihood function expressed by [25]

$$\log f(r) = -\log(\pi \det(C_{vv})) - (F_h^\# y - h)^H C_{vv}^{-1} (F_h^\# y - h) \quad (43)$$

Maximizing this log-likelihood with respect to y leads to the ML estimator given by

$$\widehat{H} = F_h F_h^\# y = P_{F_h} y \quad (44)$$

where $P_{F_h}$ denotes the orthogonal projection on the column space of $F_h (P_{F_h} = F_h (F_h^H F_h)^{-1} F_h^H$

**3.7. Kalman Channel Estimator**

Assuming that the OFDM system model used here is whose described in section I, then the state-space model of (14) and (15) allows us to use Kalman filter to adaptively track the channel gain $H_k(n)$. The algorithm is standard and is given below [21]:

1. Initialize the Kalman Filter with $x[0] = 0_{pN}$ and $\Sigma_0 = \Sigma$ where $\Sigma$ is the stationary covariance of $x[n]$ and can be computed analytically from the expression of $r_{k,l}(m)$ given in (12).

2. For each n, do the Kalman Filter update according to:

$$M_n = C\Sigma_{n-1}C^H + GG^H,$$

$$\Gamma_t = D[n]M_n D^H[n] + \sigma_\omega^2 I_N,$$

$$K_n = M_n D^H[n]\Gamma_n^{-1},$$

$$x[n] = Cx[n-1] + K_n(y[n] - D[n]Cx[n-1]),$$

$$\Sigma_n = (I_{pN} - K_n D_n)M_{n-1},$$

3. Channel estimate at instance n is

$$\hat{h}[n] = [I_N, O_N, \ldots, O_N]x[n] \tag{45}$$

Noted that the algorithm needs the information symbol $s_k[n]$, so is working in the training or decision-feedback mode.

The vector Kalman-filter algorithm gives the optimal linear estimate of the channel. Its drawback is the high complexity. Considering that the dimension of the state vector is pN which can be significantly high when there is large number of subcarriers.

One solution to reduce the complexity of the Kalman-filter channel estimator is to implement it at a per-sub-channel fashion. Consider the kth sub-channel gain $H_k(n)$. It can be modeled as a one-dimensional AR process

$$H_k(n) = -\sum_{i=1}^{p} a_i H_k(n-i) + \sigma u_k[n] \tag{46}$$

where the model parameter $a_1, \ldots, a_p$ and $\sigma$ can be computed from the correlation $r_{k,k}(0)$ of (12) according to a Yule-Walker equation. Note that these parameters do not have index k. This is because all the subcarriers have the same statistics and fit in the same AR model. This may greatly simplify the channel estimator for many components are shared by the estimator for each sub-channel. [21]

Tracking the similar process as that of the previous section, the state-space model for the kth sub-channel is obtained and expressed as:

$$x_k[n] = Cx_k[n-1] + gv_k[n] \tag{47}$$

$$y_k[n] = d_k[n]x[n] + w_k[n] \tag{48}$$

Where $x_k[n] = [H_k(n), \ldots, H_k(n-p+1)]^T$, $g = [\sigma, 0, \ldots, 0]$, $d_k = s_k[0], 0, \ldots, 0]$

$$C = \begin{bmatrix} -a_1 & -a_2 & \cdots & -a_p \\ 1 & 0 & \cdots & 0 \\ 0 & \cdots & 1 & 0 \end{bmatrix}$$

The Kalman filter for the model of $x_k[n]$ is a p-dimensional one which is much simpler. We present it as follows.

1. Initialize the Kalman Filter with $x_k[n] = 0_p$ and $\Sigma_{k,0} = \Sigma$ where $\Sigma$ is the stationary covariance of $x_k[n]$ and can be computed analytically from the expression of $r_{k,l}(m)$ given in (12).

2. For each n, do the Kalman Filter update according to:

$$M_n = C\Sigma_{n-1}C^H + gg^H,$$

$$\gamma_{k,n} = d_k[n]M_n d_k^H[n] + \sigma_\omega^2,$$

$$K_{k,n} = M_n d_k^H[n]/\gamma_{k,n},$$

$$x_k[n] = Cx_k[n-1] + K_{k,n}(y_k[n] - d_k[n]Cx_k[n-1]),$$

$$\Sigma_{,n} = (I_p - K_{k,n}d_k[n])M_{n-1},$$

3. Channel estimate at instance n is

$$\hat{H}_k[n] = [1,0,\ldots,0]x_k[n] \tag{49}$$

.

The Kalman filter estimator proposes an extended version but its drawback is when non linearity is high, the solution consists of the extension of the approximation by taking additional terms in Taylor's series

### 3.8. Subspace Approach for Blind Estimation

We consider a data model for the OFDM system model described in the section I. Let us indicate the vector $a(n) = [a_0(n), a_1(n), \ldots, a_{M-1}(n)]^T$ as the nth block of data and $s(n) = [s_{K-1}(n), [s_{K-2}(n), \ldots, [s_0(n)]^T$ as a sequence of asa sequence of the nthblock of the IDFT output and embedded cyclic prefix, where $K = M + P$ and $P$ is the length of cyclic prefix. Denoting $W \triangleq e^{j\frac{2\pi}{M}}, i = 0,1,\ldots,K-1, j = 0,1,\ldots,M-1$, the nth transmitted data sequence will be:

$$s(n) = Wa(n) \tag{50}$$

By considering N blocks of data, $= [a(n)^T, a(n-1)^T, \ldots, a(n-N+1)^T]^T$ ; s$= [s(n)^T, s(n-1)^T, \ldots, s(n-N+1)^T]^T$ and $\widetilde{W} = I_N \otimes W$ where $I_N$ is an $N \times N$ identity matrix and $\otimes$ is the Kronecker product; the received signal is expressed as:

$$r = Hs + b = H\widetilde{W}a + b \tag{51}$$

With $s = \widetilde{W}a$ and ris an $(NK - L) \times 1$ vector. Let the nthblock of received signal be denoted as $r(n) = [r_{K-1}(n), [r_{K-2}(n), \ldots, [r_0(n)]^T$ then $r = [r(n)^T, r(n-1)^T, r(n-N+2)^T, r(n-N+1)^T(1:K-L)^T]^T$ where $r(n-N+1)^T(1:K-L)^T$ is a Matlab notation standing for the first $K - L$ elements of $r(n-N+1)$. b is a noise vector that is assumed to be zero mean white

Gaussian noise with variance matrix $\sigma^2 I_{NK-L}$ and be mutually independent with the input symbol sequence. His an (NK-*L)* x *NK* matrix defined as:[6]

$$H = \begin{bmatrix} h_0 & \cdots & h_L & 0 & \cdots & & 0 \\ 0 & h_0 & \cdots & h_L & 0 & \cdots & 0 \\ \vdots & \cdots & \ddots & \cdots & & \ddots & \vdots \\ 0 & \cdots \cdots & 0 & h_0 & \cdots & & h_L \end{bmatrix} \tag{52}$$

Denote $A \triangleq H\widetilde{W}$n then (51) becomes

$$r = Aa + b \tag{53}$$

Since the matrix *A*should be full column rank for the channel to be identified, we give a sufficient condition for full rank requirement; this will be verified under the assumption $L \leq PN$ [6].

In fact it is the inserted prefix that makes matrix *A*a "tall" matrix and be possible to be full column rank. The full column rank condition can always be satisfied aslong as*N* is chosen to be large enough.

The identification problem is based on the $(NK - L) \times (NK - L)$ autocorrelation matrix $R_r$of the measurement where $R_r = E\{rr^H\}$. Considering the expression given in (53), this autocorrelation matrix will be:

$$R_r = AR_aA^H + \sigma^2 I \tag{54}$$

If matrix **A** is fullcolumn rank and the autocorrelation of input $R_a$, is alsofull rank, then $range(A) = range(AR_aA^H)$. Let us define the noise subspace for $R_r$, to be the subspace generated by *PN* - L eigenvectors corresponding to the smallest eigenvalue, and let $G = [G_1, \ldots, G_{PN-L}]$be the matrix containing those eigenvectors. Then, $G$ spans the null space of $AR_aA^H$ and is orthogonal to its range space:

$$G_i^H A = 0 \quad i = 1, \ldots, PN - L \tag{55}$$

For determining the identifiability of the channel, the equation above will be solved in the least square sense. Let us define $\xi_k$ as:

$$\xi_k = \begin{bmatrix} G_{k,0} & \cdots & G_{k,J} & 0 & \cdots & & 0 \\ 0 & G_{k,0} & \cdots & G_{k,J} & 0 & \cdots & 0 \\ \vdots & \cdots & \ddots & \cdots & & \ddots & \vdots \\ 0 & \cdots \cdots & 0 & & G_{k,0} & \cdots & G_{k,J} \end{bmatrix} \tag{56}$$

Where $J = \text{KN} - L - 1$. Next, $G_k^T H = h^T \xi_k$. This leads to the minimization problem:

$$\hat{h} = \arg\min \sum_{k=1}^{PN-L} \hat{G}_k^H H \widetilde{W} \widetilde{W}^H H^H \hat{G}_k \tag{57}$$

$$= \arg\min \sum_{k=1}^{PN-L} h^H \hat{\xi}_k \widetilde{W} \widetilde{W}^H \hat{\xi}_k^H h$$

$$= \arg\min \sum_{k=1}^{PN-L} h^H \Psi h$$

Where $\Psi = \sum_{k=1}^{PN-L} \hat{\xi}_k \widetilde{W} \widetilde{W}^H \hat{\xi}_k^H$. Minimization is subject to the constraint $\|h\| = 1$ Therefore, $\hat{h}$ is given by the eigenvector corresponding to the smallest eigen value of $\Psi$. [6]

## 4. SIMULATIONS RESULTS

In this section, we will evaluate the performance of each method of channel estimation presented in the previous sections in term of Bit Error Rate BER and Mean Square Error MSE, Root Mean Square Error RMSE and some figures will display the estimated samples behind the transmitted samples.

For all the algorithms considered, the channel adopted is a Rayleigh fading one with either 16-QAM or BPSK modulations

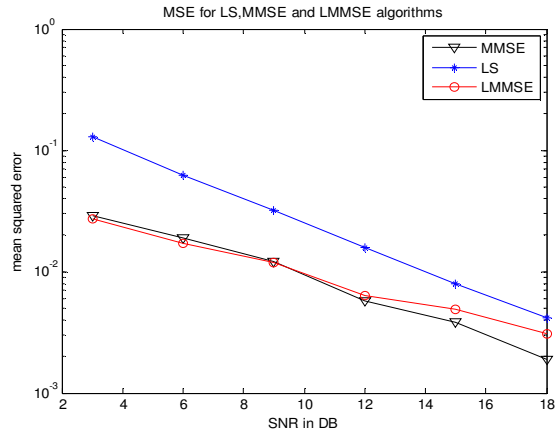

Figure 1 : MSE vs SNR with a 16 QAM modulation

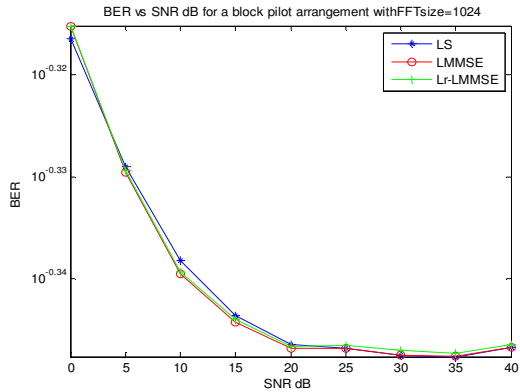

Figure 2 : SNR vs BER with a 16 QAM modulation

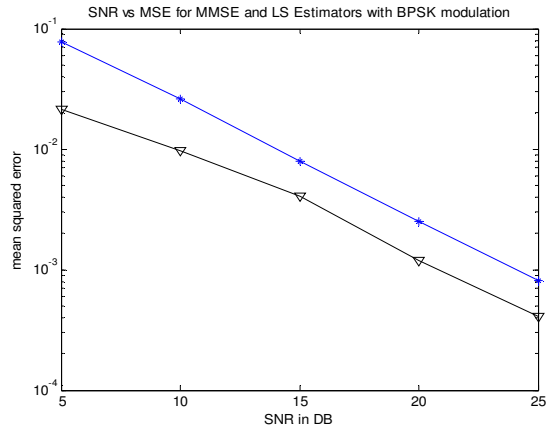

Figure 3 : SNR vs MSE with BPSK modulation

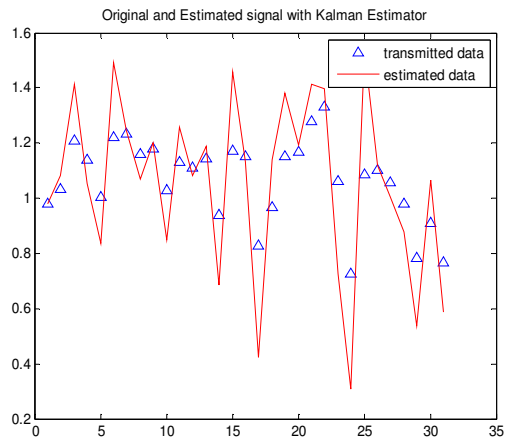

Figure 4: Kalman Filter Estimator

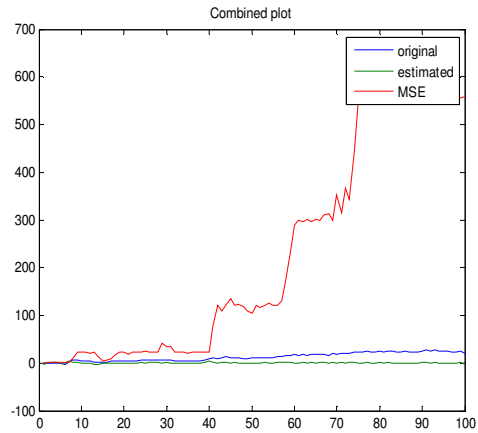

Figure 5: MSE evaluation for an Extended Kalman Filter Estimator

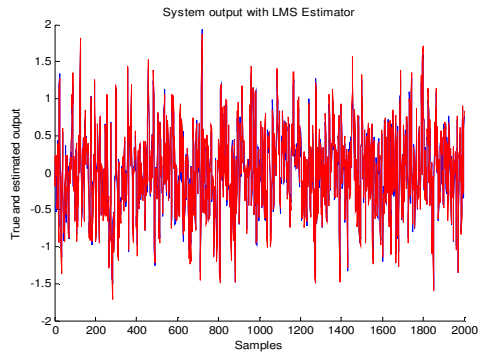

Figure 6: LMS Estimation display of transmitted and received symbols

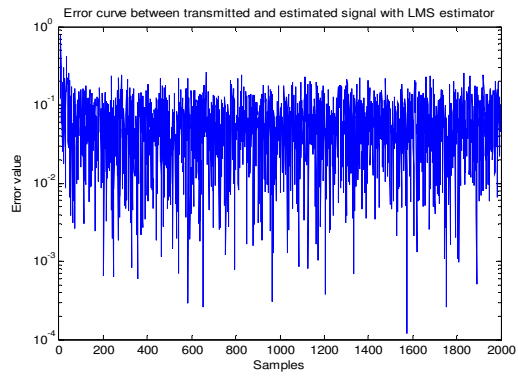

Figure 7: LMS Error Curve between transmitted and estimated signals

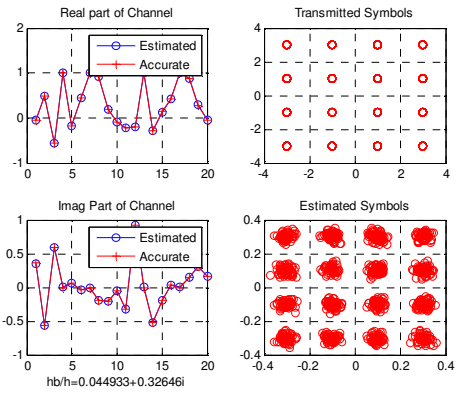

Figure 8: Subspace method for blind estimation with a 16 QAM modulation

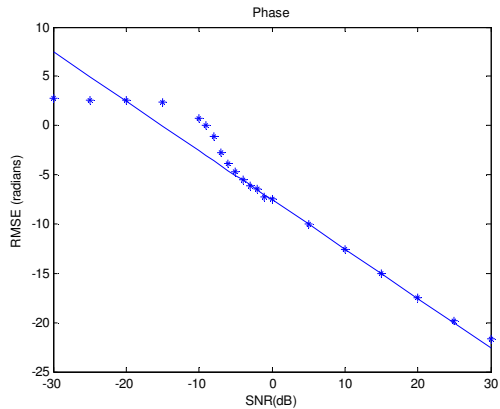

Figure 9: RMSE vs SNR for a ML Estimator of the phase

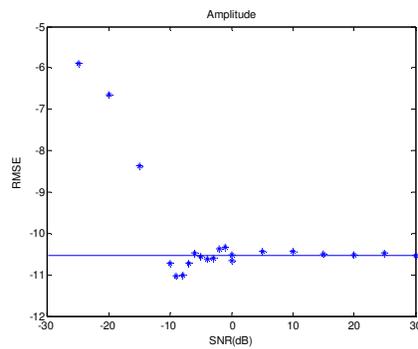

Figure 10: RMSE vs SNR for a ML Estimator of the amplitude

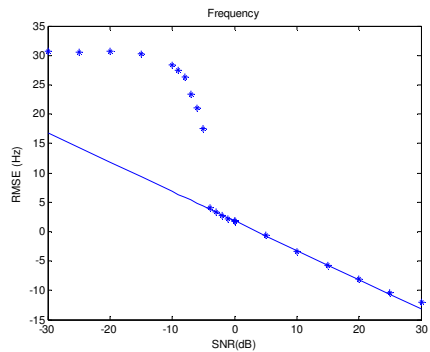

Figure 11: RMSE vs SNR for a ML Estimator of the frequency

## 4. CONCLUSIONS

In this paper, the main methods and algorithms of channel estimation for a time varying channel was presented and some of them performed. Theses methods can be structured in two major categories: one is dedicated for slowly time varying channels and the one for the fast time varying channels.

The performance of the channel was evaluated in term of bit error rate and mean square error versus SNR and for some of them the root mean squared error. Some of results show the difference between the transmitted and received data

The results illustrate that the Least Square is the basis of the major channel estimation algorithms and gives usually a high BER comparatively with MMSE and LMMSE algorithms. The ML and blind estimators are also robust estimators even if they remain more complex.

As a prospective of our work is that we can extend all these methods to the case of MIMO OFDM channel estimation.

**Authors**

**PR. RIDHA BOUALLEGUE**

Received the Ph.D degrees in electronic engineering from the National Engineering School of Tunis. In Mars 2003, he received the Hd.R degrees in multiuser detection in wireless communications. From September 1990.He was a graduate Professor in the higher school of communications of Tunis (SUP'COM), he has taught courses in communications and electronics. From 2005 to 2008, he was the Director of the National engineering school of Sousse. In 2006, he was a member of the national committee of science technology. Since 2005, he was the laboratory research in telecommunication Director's at SUP'COM.From 2005, he served as a member of the scientific committee of validation of thesis and Hd.R in the higher engineering school of Tunis. His recent research interests focus on mobile and wireless communications, OFDM, OFDMA, Long Term Evolution (LTE) Systems. He's interested also in space-time processing for wireless systemsand CDMA systems.

**ZAIER AIDA**

Received the B.S. degree in 2005 from National Engineering School of Gabes, Tunisia, and M.S. degree in 2006 from Polytechnic School of Sophia Antipolis of Nice Frrance. Her Research interests focus on channel estimation and synchronization of OFDM and  MIMO-OFDM channels under very high mobility conditions.